\documentclass[%reprint,
4pt,eqsecnum,
superscriptaddress,
preprint,
 amsmath,amssymb,
 aps,
]{revtex4-1}

\usepackage{graphicx}% Include figure files
\usepackage{dcolumn}% Align table columns on decimal point
\usepackage{hyperref}
\usepackage{bm}% bold math
\usepackage{enumerate}
\usepackage{lipsum}
\usepackage{epstopdf}
\usepackage{float}
\usepackage[caption = false]{subfig}
\usepackage{tabularx}
\begin{document}

\preprint{APS/123-QED}

\title{Interaction of atom with non-paraxial Laguerre-Gaussian beam: Forming superposition of vortex states in Bose-Einstein condensates}
\author{Anal Bhowmik}
\affiliation{Department of Physics, Indian Institute of Technology Kharagpur, Kharagpur-721302, India.}
\author{Pradip Kumar Mondal}
%\email{pradip3@phy.iitkgp.ernet.in}
\affiliation{Department of Applied Science, Haldia Institute of Technology, Haldia-721657, India.}
\author{Sonjoy Majumder}
\email{sonjoym@phy.iitkgp.ernet.in}
\affiliation{Department of Physics, Indian Institute of Technology Kharagpur, Kharagpur-721302, India.}
\author{Bimalendu Deb}%
% \email{msbd@iacs.res.in}
\affiliation{Department of Materials Science, Indian Association for the Cultivation of Science, Jadavpur, Kolkata 700032, India.}
\date{\today}


%\affiliation{
%\affiliation{
 %Third institution, the second for Charlie Author
%\affiliation{%



\begin{abstract}
The exchange of orbital angular momentum (OAM) between paraxial optical vortex and a Bose-Einstein condensate (BEC) of atomic gases is well known. In this paper, we develop a theory for the microscopic interaction between matter and an optical vortex beyond paraxial approximation. We show  how superposition of vortex states of BEC can be created  with a focused optical vortex.  Since, the polarization or spin angular momentum (SAM) of the optical field is coupled with OAM of the  field, in this case, these
angular momenta can be transferred to the internal electronic and external center-of-mass (c.m.)
motion of atoms provided both the motions are coupled. We propose a scheme for producing the superposition of matter-wave vortices using  Gaussian and a focused Laguerre-Gaussian (LG) beam. We  study  how two-photon Rabi frequencies of stimulated  
Raman transitions   vary with  focusing angles for different combinations of OAM and SAM of optical  states.     We demonstrate the formation of   vortex-antivortex structure and discuss  interference of three vortex states in a BEC.

\end{abstract}

\maketitle

\section{INTRODUCTION}
Recognition of orbital angular momentum (OAM) of light  has evoked a lot of activities in different branches of physics over last two decades. Spin angular momentum (SAM) is carried by the polarization of light while OAM is by helical phase front. Being an extrinsic property, OAM generally affects the c.m. motion of an atom, whereas,  SAM of field determines the selection rules of electronic transitions. In our recent work \cite{Mondal2014}, we have shown that optical
OAM can be transferred to electronic motion  via
quantized c.m. motion of ultracold atoms within paraxial approximation. For focused optical vortex beam, parxial approximation breaks down and non-paraxial effects \cite{Allen1992} become important.   New realm of physics can be explored for atoms or molecules interacting with non-paraxial (focused) optical vortex where the SAM and the OAM are no longer conserved separately but the total angular momentum (OAM+SAM) is conserved in interaction with an atom or a molecule \cite{Marrucci2006,Zhao2007}. The interesting feature of focused optical vortex is that the OAM of light can be  transferred to the electronic motion or the SAM of light can affect the c.m. motion of an atom even at dipole approximation  level unlike that in the case of   paraxial approximation. Considering direct coupling of field OAM with the internal motion of atoms,  many applications are proposed  in literature, such as second-harmonic generation in nonlinear optics \cite{Romero2002}, new selection rules in photoionization \cite{Picon2010,Picon2010_2,schmi_12}, strong dichroism effect \cite{veenedaal_07,Mondal2015}, charge-current generation in atomic systems \cite{koksal_12}, the suppression of parasitic light shifts in the field of quantum information and metrology experiments with single atoms or ions \cite{schmi_12}, new selection rules in off-axis photoexcitation \cite{afanasev_13},  etc.   The non-paraxial vortex beams have   applications in different fields of research such as,  quantum information processing \cite{Beugnon2007}, trapping of atoms \cite{Chu1986} or microparticles \cite{Ashkin1986} in optical twizers, cell biology \cite{Mehta1999} etc.

Here we develop a theory   for the interaction of non-paraxial vortex beam with an atom and apply this to the creation of superposition of matter wave vortices in an atomic Bose Einstein condensate (BEC). We show the possibility to  create  multiple quantized circulations of BEC using single focused  optical vortex  pulse, unlike that in earlier works \cite{Liu2006,Wen2008,Wen2013}  where multiple optical vortices were used.    To transfer the OAM from the light to the c.m. motion of matter, the wavelength of the matter wave has to be  large enough to feel the intensity distribution of the optical vortex beam. So, this theory will be applicable to cold atoms. Since the spread of wavefunction of cold single trapped atom is very narrow, transfer mechanism is appreciable for large number of cold atoms, like,  BEC. The main question we address in this paper is about the sharing mechanism of the total angular momentum of a focused optical vortex  between the external c.m. and internal electronic motions of an atom.  We show that there are  three possible ways of distributing the total field angular momentum between c.m. and electronic motions. We call them as angular momentum channels (AMC) of interaction. The atoms interact with the LG beam via different AMCs having probabilities that depend on corresponding transition strengths and focusing angles.

 The formalism  of corresponding interaction is developed in Section II.  Section III describes numerical calculations of a proposed method of creation of superposition of BEC vortex states using non-paraxial LG beam. Section IV discusses some examples of superposition of BEC vortex states,   like vortex-antivortex pair, which can be created by our proposed method giving simulated interference patterns. Finally, in Section V, we make some concluding remarks.

\section{THEORY}
The focused non-paraxial beam considered here is produced from a circularly polarized paraxial pulse with OAM by passing  it through a lens with high numerical aperture (NA). The consequent spin-orbit coupling of light is based on Debye-Wolf theory  \cite{Richards1959, Boivin1965}, where an
incident collimated LG beam is decomposed into a
superposition of plane waves having an infinite number of
spatial harmonics. In a non-paraxial
beam, the total angular momentum is a good quantum
number. In the rest of the paper, whenever we mention about SAM or OAM, it should
be understood that we mean the corresponding angular momentum of the paraxial LG beam before passing
through the lens. We consider that the focused $LG^l_p$ beam ($l$ is OAM of light beam \cite{Allen1992} and $p$ is radial node of Laguerre polynomial) interacts with
cold atoms whose de Broglie wavelength is large enough
to feel the intensity variation of the focused beam.   For non-paraxial circularly polarized $ LG_0^l $ beam, the x, y, z-polarized component of the electric field \cite{Zhao2007,Monteiro2009,Iketaki2007} in the laboratory coordinate system can be expressed   as

\begin{equation}
{E_x}(r^\prime,\phi^{\prime},z^\prime)=(-i)^{l+1}E_0(e^{il\phi ^\prime}I_0^{(l)}+e^{i(l+2\beta)\phi ^\prime}I_{2\beta}^{(l)}),
\end{equation}
\begin{equation}
{E_y}(r^\prime,\phi^{\prime},z^\prime)=\beta(-i)^{l}E_0(e^{il\phi ^\prime}I_0^{(l)}-e^{i(l+2\beta)\phi ^\prime}I_{2\beta}^{(l)}),
\end{equation}
\begin{equation}
{E_z}(r^\prime,\phi^{\prime},z^\prime)=-2\beta(-i)^{l}E_0e^{i(l+\beta)\phi ^\prime}I_{\beta}^{(l)},
\end{equation}
where $\beta$ is the polarization of light incident on the lens. Here, we consider only circular polarization with  $\beta = \pm 1$.   The amplitude of the focused electric field is $E_0=\dfrac{\pi f}{\lambda} T_{o} E_{inc}$, where we have assumed $T_{o}$ is the objective transmission amplitude, $E_{inc}$ is the amplitude of incident electric field and $f$ is the focal length related with $r^\prime$ by $r^\prime=f \sin\theta$ (Abbe sine condition). The coefficients $I_m^{(l)}$, where $m$ takes the values 0, $\pm1$, $\pm2$   in the above expressions depend on focusing angle ($\theta_{max}$) by \cite{Zhao2007}

\begin{eqnarray}
I_m^{(l)}(r _\bot ^\prime ,z ^\prime)&=&\int_0^{\theta_{max}}d\theta\left({\dfrac{\sqrt{2}r_\bot^\prime }{w_0 \sin\theta}}\right)^{\lvert l \rvert}{(\sin\theta)}^{\lvert l \rvert +1} \nonumber \\
&\sqrt{\cos\theta}& g_{\lvert m \rvert}(\theta) J_{l+m}(kr_\bot^\prime \sin\theta)e^{ikz^\prime \cos\theta},
\end{eqnarray} 
where $r_\bot^\prime$ is the projection of \textbf{r$^\prime$} on the $xy$ plane, $w_0$ is the waist of the paraxial beam and $J_{l+m}(kr_\bot^\prime \sin\theta)$ is cylindrical Bessel function. The angular functions are  $g_0 (\theta)=1+\cos\theta$, $g_1 (\theta)=\sin\theta$, $g_2 (\theta)=1-\cos\theta$.

We consider  the  above field  interacts with simplest atomic system with a core of total charge $+e$ and mass $m_c$ and a valance electron of  charge $-e$ and mass $m_e$.  The
c.m. coordinate with respect to laboratory coordinate system  is $\textbf{R}=(m_e \textbf{r}_e + m_c \textbf{r}_c)/m_t , m_t=m_e+m_c$ being the total mass and their relative (internal) coordinate is given by $\textbf{r}=\textbf{r}_e -\textbf{r}_c$ \cite{Mondal2014}. Here $\textbf{r}_e$ and $\textbf{r}_c$ are the coordinates of the valance electron and the center of atom,  respectively, with respect to laboratory coordinate system.

The atomic system is trapped in a harmonic potential and  the atomic state can be written as a product of the c.m. wave function and electronic wave function $\Upsilon(\textbf{R},\textbf{r})=\Psi_R(\textbf{R})\psi(\textbf{r})$.  The c.m. wave function $\Psi_R(\textbf{R})$ depends on the external harmonic trapping potential.  The internal electronic wave function $\psi(\textbf{r})$ can be considered as a highly correlated coupled-cluster state    \cite{Mondal2013}. The interaction Hamiltonian $H_{int}$ is derived using the Power-Zienau-Wooley (PZW) scheme \cite{Babiker2002}. Since $\lvert \textbf{r}\rvert \ll\lvert \textbf{R}\rvert $,  we can use the Taylor's expansion for the electric field about \textbf{R}. For circularly polarized light, with OAM=+1, the electric dipole interaction Hamiltonian becomes [see Appendix, Eq. (A.8)]
 
\begin{eqnarray}
H_{int}^{ l=+1,\beta=\pm 1}&=&  e\dfrac{m_c}{m_t} r \sqrt{\dfrac{8\pi}{3}}\Bigl[-I_0^{(1)}(R_\bot,Z)e^{i\Phi}\epsilon_{\pm 1}Y_1^{\pm 1}(\boldsymbol{\hat{\textbf{r}}})\nonumber \\
&-& I_{\pm 2}^{(1)}(R_\bot,Z)e^{i(1\pm 2)\Phi}\epsilon_{\mp 1}Y_1^{\mp 1}(\boldsymbol{\hat{\textbf{r}}})\, \nonumber\\
&\pm & \sqrt{2} i I_{\pm 1} ^{(1)}(R_\bot,Z)e^{i(1 \pm 1)\Phi}\epsilon_{=0}Y_1^{0}{(\boldsymbol{\hat{\textbf{r}}}})\Bigr].
\end{eqnarray}
Here  $H_{int}$ depends  mainly on two parameters, i.e., orbital ($l$) and spin ($\beta$) angular momentum of light. The electric dipole transition selection rule is $\Delta l_e=\pm 1$, $\Delta m_l=0,\pm 1$. Here $l_e$ and $m_l$ are the electronic orbital angular momentum and its projection along the direction of propagation of the light i.e. laboratory z-axis. In interaction with paraxial beam,  any one of the above conditions   for $\Delta m_l$ is satisfied, depending on the polarization of light ($\beta = 0, \pm 1$) and we have only one AMC of interaction. But in interaction with non-paraxial light all the possibilities of $\Delta m_l$ open up for $\beta = 1$ or -1. These generate three possible electronic  hyperfine sublevels, as discussed later, in the atoms of BEC as seen from Eq. (5).  But total angular momentum has to be conserved. Therefore, as seen from the equation, we get three possible orbital angular momentum states of the c.m. of the atoms  corresponding to the above electronic states.

Let us now discuss each term
of Eq. (5) to understand how the SAM and OAM of the
incident paraxial beam  are shared between the electronic and c.m. motion of the atom.  First term of this equation represents the paraxial-term i.e.,  the OAM of light interacts with the c.m. motion and the polarization of light interacts with the electronic motion of the atom \cite{Romero2002,Jauregui2004,Picon2010}. But  the second and third terms of this equation imply that the polarization of the light can also affect the external motion of c.m. of the atoms. The three terms sequentially represents three channels refer as AMC-1, AMC-2 and AMC-3, respectively. With the increase of the focusing,
light changes its vector properties and the possibilities of
conversion of SAM to OAM increases  \cite{Marrucci2006, Zhao2007}. This implies  that AMC-2 and AMC-3 will become more significant with
increasing the focusing angle by changing the NA of the lens. One part of the total angular momentum (TAM) goes to the c.m. and creates the vorticity of the matter-wave. If any part of TAM goes
to the electron, it generates electronic transitions satisfied by the electromagnetic selection rules. Therefore, the dipole
transition matrix element between two states ($\rvert \Upsilon _i \rangle$ and $\rvert \Upsilon _f \rangle$) of the system is given by 
\begin{widetext}
\begin{align}
 M_{i \rightarrow f}^d  =  \langle \Upsilon _f \lvert H_{int}^{l=+1, \beta=\pm 1} \rvert \Upsilon _i \rangle \, \nonumber 
 =  e\dfrac{m_c}{m_t}  \sqrt{\dfrac{8\pi}{3}}\Bigl[-\epsilon_{\pm 1}\langle \Psi _{R,f} \lvert I_0^{(1)}(R_\bot,Z)e^{i\Phi} \rvert \Psi _{R,i} \rangle \langle \psi _f \lvert r Y_1^{\pm 1}(\boldsymbol{\hat{\textbf{r}}})\rvert \psi _i \rangle \,\nonumber \\
- \epsilon_{\mp 1} \langle \Psi _{R,f} \lvert I_{\pm 2}^{(1)}(R_\bot,Z)e^{i(1\pm 2)\Phi}\rvert \Psi _{R,i} \rangle          \langle \psi _f \lvert r Y_1^{\mp 1}(\boldsymbol{\hat{\textbf{r}}})\rvert \psi _i \rangle \, \nonumber\\ 
\pm   \sqrt{2} i \epsilon_{0} \langle \Psi _{R,f} \lvert  I_{\pm 1}^{(1)}(R_\bot,Z)e^{i(1\pm 1)\Phi} \rvert \Psi _{R,i} \rangle \langle \psi _f \lvert r Y_1^{0}{(\boldsymbol{\hat{\textbf{r}}}})\rvert \psi _i \rangle \Bigr].
\end{align}
\end{widetext}
The three terms in Eq. (6) correspond to vorticities $l$, $l\pm 2$, $l\pm 1$ respectively, as seen from the first factors.  Second factors correspond to the transition matrix elements for electrons. These factors are numerically evaluated (see Section IV) after estimating the wavefunctions of c.m. and electronic states of the system.
In the next section, we  study two-photon stimulated Raman
transition using a focused LG beam and predict 
interesting effects.

\section{CREATION OF SUPERPOSITION OF BEC VORTEX STATES}
Generation of quantized vortices in a BEC using optical vortex has become important due to the experimental endeavors \cite{Andersen2006,Wright2008}  over last decade. The coherent superpositions of  vortices of different circulation quantum numbers, especially vortex-antivortex cases \cite{Wright2009, Wright2008}, yield interesting interference effects with potential applications \cite{Brachmann2011, Quinteiro2014}, such as manipulating the chirality of twisted metal nano-structures \cite{Toyoda2012}.    Creation of matter-wave vortex states from a non-rotating BEC by two-photon Raman transition method under  paraxial LG and Gaussian (G) pulse is well discussed in literature \cite{nandi_04,Andersen2006, simula_08, song_09, wright_08, wright_09, jaouadi_10, gullo_10, tasgin_11, lembessis_11, brachmann_11, kanamoto_11, ramanathan_11, robb_12, okulov_12, beattie_13}.  In these studies, matter-wave vortex is shown to  acquire vorticity  equal to the winding number of the LG beam.

 We consider a focused LG beam is interacting with  a  non-rotating  $ ^{23} $Na BEC,   prepared in $\lvert \psi _i \rangle=\lvert 3S_{\frac{1}{2}}, F=1, m_f =-1 \rangle$ state  in a harmonic potential. The LG pulse  induces dipole transitions in atoms as given in Eq. (6). The final state will have three different  hyperfine sublevels shown in Fig. 1. To bring back the matter in the initial state using  two-photon stimulated Raman transition, we require three simultaneous co-propagating Gaussian pulses with suitable frequencies  and polarizations  to be shined in the same direction of LG field as Shown in Fig. 2.  Because of co-propagation of the Gaussian pulses with LG field, net transfer of linear momentum to the atoms is zero.  Two-photon transitions of matter state through the three hyperfine sublevels of excited state can be defined as three channels discussed below. This procedure yields the possibility of three vorticities  in the BEC  and  creates the superposition of vortices at the initial hyperfine sublevel. Since the interference pattern of the superposition will depend on the populations of the vortex states,   the Rabi frequencies corresponding to these  two-photon transitions are important to quantify.

\begin{figure}[h]

\centering
\includegraphics[trim={0.5cm 0.5cm 1cm 1cm},width=5cm]{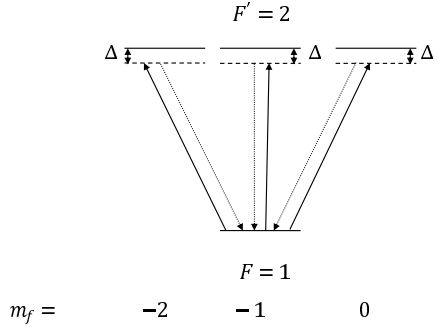}
\caption{Energy
level scheme of the two-photon transitions. The atomic states
show the  $^{23}$Na hyperfine states. Atoms  are initially in $\lvert  3s_{\frac{1}{2}} F=1, m_f =-1 \rangle$. $\Delta$ represents two-photon detuning.}
\end{figure}

\begin{figure}[h]
\centering

\includegraphics[trim={0.2cm 3cm 1.5cm 1.5cm},width=8cm]{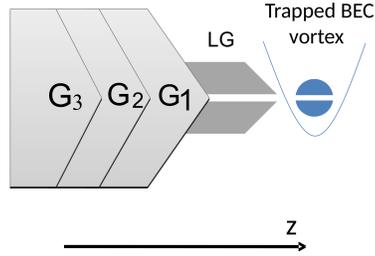}   
\caption{Single LG and three gaussian (G$_1$, G$_2$, G$_3$) pulses are applied to BEC.}
\end{figure}

 For  axial confinement of the trap,  the quantum state of the
condensate can be described by a wave function $\Psi (X,Y,t)$ in
two dimensions. In the zero-temperature limit, the dynamics
of the  weakly interacting BEC  is described  by the 
Gross-Pitaevskii equation  in cylindrical coordinate system. Let us consider   a non-paraxial LG beam, produced from a paraxial LG field with OAM=+1 and SAM=+1, and followed by gaussian beams incident on BEC.  As a result, a superposed vortex state with vorticity $\kappa= 1,2,3$ will be created. In general, the three different macroscopic vortices  with vorticities $l$, $l+\beta$, $l+2\beta$ (originated from OAM=$l$ and SAM=$\beta$) superpose  with arbitrary proportion and this superposition can be expressed as \cite{Liu2006}

\begin{equation}
\Psi(R,\Phi, t)= f(R) e^ {-i\mu t}(\alpha_1 e^{il\Phi} + \alpha_2 e^{i(l+2\beta)\Phi} + \alpha_3 e^{i(l+\beta)\Phi}),
\end{equation}
 where $R^2=(X^2+Y^2)$, $\mu$ is chemical potential of the system.  $\alpha_1, \alpha_2 $ and $\alpha_3$ are  constants, depended on the strengths of two-photon transitions corresponding to different vortex channels  with $ |\alpha_1 |^2 +| \alpha_2 |^2 + | \alpha_3|^2=1$ .  Interestingly, for the combination of (OAM, SAM)=(1, -1) or (-1, 1) of incident field , we get superposition of vortex states of BEC in the trap with $\kappa=0,1,-1$. Therefore, this turns out to be an unique approach to create superposed state of vortex-antivortex from a single LG beam.

\section{NUMERICAL RESULTS AND INTERPRETATION}

We start with single photon scattering by trapped atoms as expressed in Eq. (6). For numerical calculations, we choose the characteristics of the experimental trap as given in Ref \cite{Andersen2006} with asymmetry parameter  $\lambda _{tr} =\omega _Z /\omega _\bot =2$ and the axial frequency $\omega _Z /2\pi =40$ Hz. The characteristic length and $s$-wave scattering length are  $a _\bot =4.673$  $\mu$m and $a=2.75$ nm,  respectively. The intensity of the paraxial LG beam, which has been focused, is $I =10 $ mW m$^{-2}$ and its waist $w _0 =10 ^{-4}$ m. We now  numerically evaluate the Rabi frequencies of dipole transitions  considering the  Eq. (6) where the c.m. and electronic motions are coupled. Let us consider a left circularly polarized paraxial LG beam (means SAM=+1) with OAM=+1 transforms into non-paraxial LG  beam and interacts with  a non-rotating BEC of $10^5$ ${^{23}}$Na atoms in an anisotropic harmonic trap. The axes of the beam and the trap are  along the z axis of the laboratory frame.

 In Eq. (6), $\langle \psi _f \rvert r Y_1 ^{0,\pm 1}(\boldsymbol \hat{\textbf{r}})\lvert \psi _i \rangle$  is the electronic portion of the dipole transition due to the interaction  with LG beam, but interestingly depends on,   the vorticity of c.m. motion of BEC. The vorticity of excited  state with hyperfine sublevels $m_f=0,-1,-2$ will be $l$, $l+1$, $l+2$ for SAM=+1 of paraxial field.  

FIG. 3 shows that  Rabi frequencies of different transitions with LG field of OAM=+1 and SAM=-1. These  results show that the values of matrix elements of two-photon transitions increase significantly with focusing angles.   Note that, $\lvert F=1, m_f =-1 \rangle$ $\rightarrow$  $\lvert F=2, m_f =0 \rangle $  and $\lvert F=1, m_f =-1 \rangle$ $\rightarrow$      $\lvert F=2, m_f =-1 \rangle $ transitions are negligible under paraxial approximation. Here in non-paraxial case, we notice that these transitions are non-negligible and become significant with high focusing angles.    The finiteness of these two transitions at  small focusing angle ($\approx 10^\circ$)  may be due to the inclusion of diffraction feature during the conversion of paraxial to non-paraxial beam. Interestingly, the relative  strength of these two weak transitions changes  as we change the  focusing angle. The similar  features for other combinations of OAM and SAM of light are also observed and will be discussed below.

\begin{figure}[h]

  \centering

\includegraphics[width=200mm]{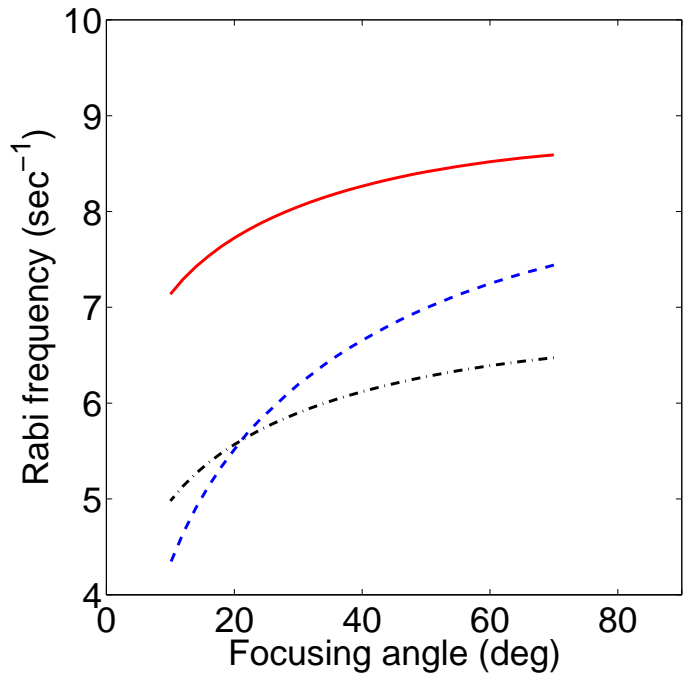}

\caption{Variations of dipole  Rabi frequency (in sec$^{-1}$) with focusing angles (in degrees$^\circ$) are plotted on a
semi-log scale. Red solid line refers to elctronic transition $\lvert F=1, m_f =-1 \rangle$  to   $\lvert F^\prime=2, m_f =-2 \rangle $, Blue dashed line is for $\lvert F=1, m_f =-1 \rangle$  to   $\lvert F^\prime=2, m_f =0 \rangle $, and dotted line represents $\lvert F=1, m_f =-1 \rangle$ to $\lvert F^\prime=2, m_f =-1 \rangle $.}

\end{figure}

To calculate the two-photon Rabi frequencies, we consider that  co-propagating LG and a set of Gaussian (G) beams interact with the trapped BEC as shown in FIG. 2.    Let us consider the atoms which will take part
in the two-photon transitions will reach final electronic
state $\lvert 3S_{\frac{1}{2}} F=1, m_f =-1 \rangle$. It means the final internal atomic state will be same as the initial one  which is low field seeking.  The frequency difference between the
two kinds of pulses, $\delta \nu_r$,  is
the recoil energy. Here G beam is
detuned from the D1 line  by $\Delta=-1.5$ GHz
 ($\approx -150$ linewidths,       enough to prevent any significant spontaneous photon scattering). We
apply  LG/G beams to the trapped atoms
and look for the superposition of vortex states.   TABLE 1. shows the results of two-photon Raman  transitions with three channels going through three intermediate states,  $\Omega_1= \lvert F^\prime=2, m_f =-2 \rangle$, $\Omega_2= \lvert F^\prime=2, m_f =0 \rangle$ and $\Omega_3= \lvert F^\prime=2, m_f =-1 \rangle$. As expected from the single LG photon absorption,  $\Omega_1$  is always greater than $\Omega_2$ and $\Omega_3$. But crossing of  amplitudes of $\Omega_2$ and $\Omega_3$ happens at $\approx 30^\circ$ unlike single photon transition (happened at $\approx 20^\circ$). The point to be noted here is that   $\Omega_1$ and $\Omega_2$ correspond to vorticities 1 and -1, respectively. At high focusing angle, the ratio between the strength of $\Omega_1$ and $\Omega_2$ decreases and interference pattern will clearly be visible as a superposition of   vortex and anti-vortex  as shown in FIG. 4.   

\begin{table}[h]
\caption{Magnitude of Rabi frequencies (in MHz) of two-photon Raman transitions for different focusing angles of incident beam of OAM=+1, SAM=-1. $\kappa$ is the final vorticity of atoms in BEC. } % title of Table
\centering % used for centering table
\begin{tabular}{c |c|c | c} % centered columns (4 columns)
\hline\hline %inserts double horizontal lines
Focusing angle & $\Omega_1 (\kappa=1)$ & $\Omega_2 (\kappa=-1)$ & $\Omega_3 (\kappa=0)$ \\ [0.2ex] % inserts table
%heading
\hline % inserts single horizontal line
70$^\circ$ & $456.50$ & $13.17 $ & $2.47$ \\ % inserting body of the table
60$^\circ$ & $386.46$ & $8.46$ & $2.04$ \\
50$^\circ$ & $303.41$ & $4.69$ & $1.56$ \\
40$^\circ$ & $215.14$ & $2.15$ & $1.09$ \\
30$^\circ$ & $131.34$ & $0.75$ & $0.65$ \\ 
20$^\circ$ & $61.94$ & $0.16 $ & $0.31$ \\
10$^\circ$ & $15.95$ & $0.01 $ & $0.08 $ \\[0.2ex]
 % [1ex] adds vertical space
\hline %inserts single line
\end{tabular}
\label{table:nonlin} % is used to refer this table in the text
\end{table}

\begin{figure*}[!h]
\subfloat[]{\includegraphics[trim = 1cm 1.5cm 1cm 2cm, scale=.22]{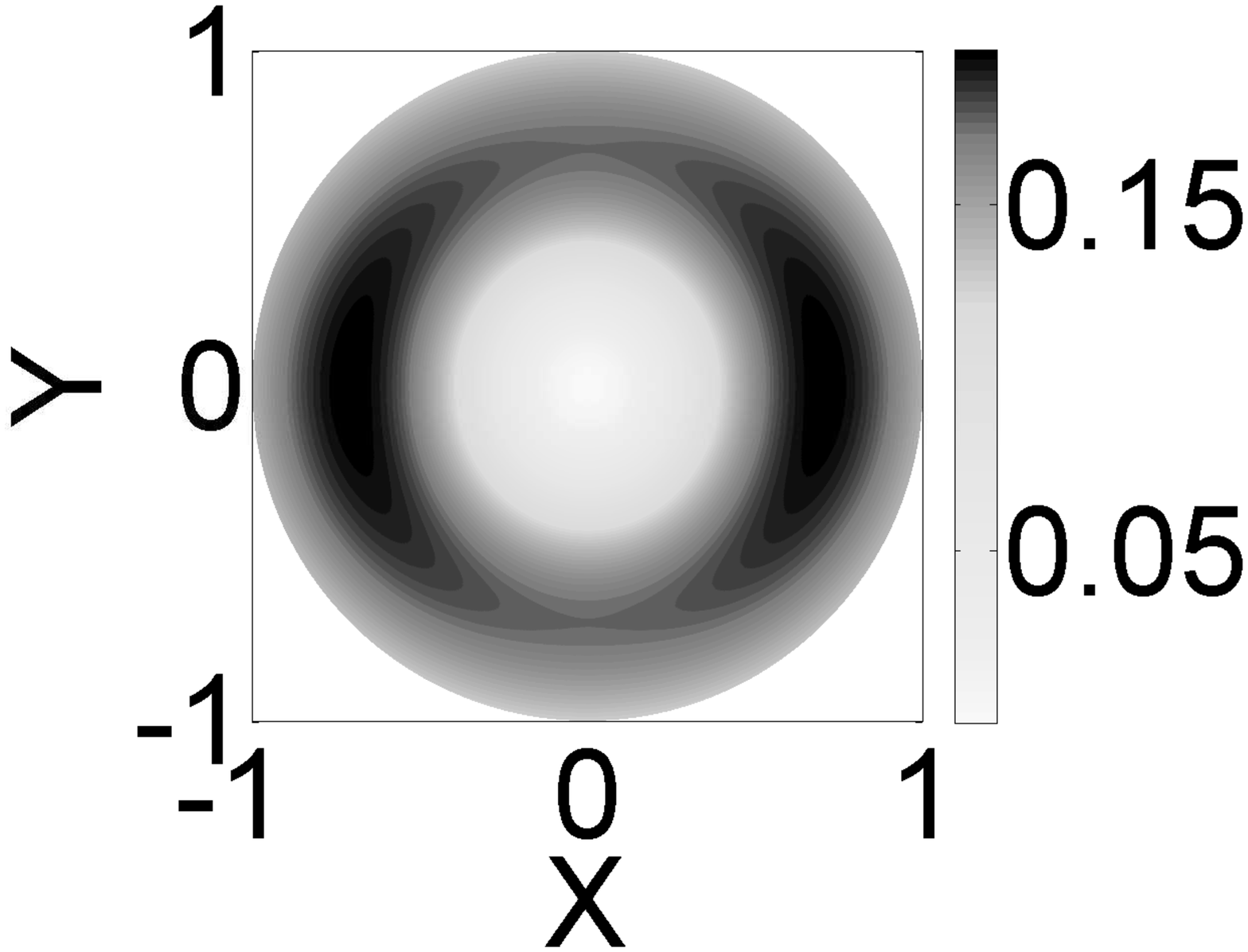}}
\subfloat[]{\includegraphics[trim = 1cm 1.5cm 1cm 0cm, scale=.22]{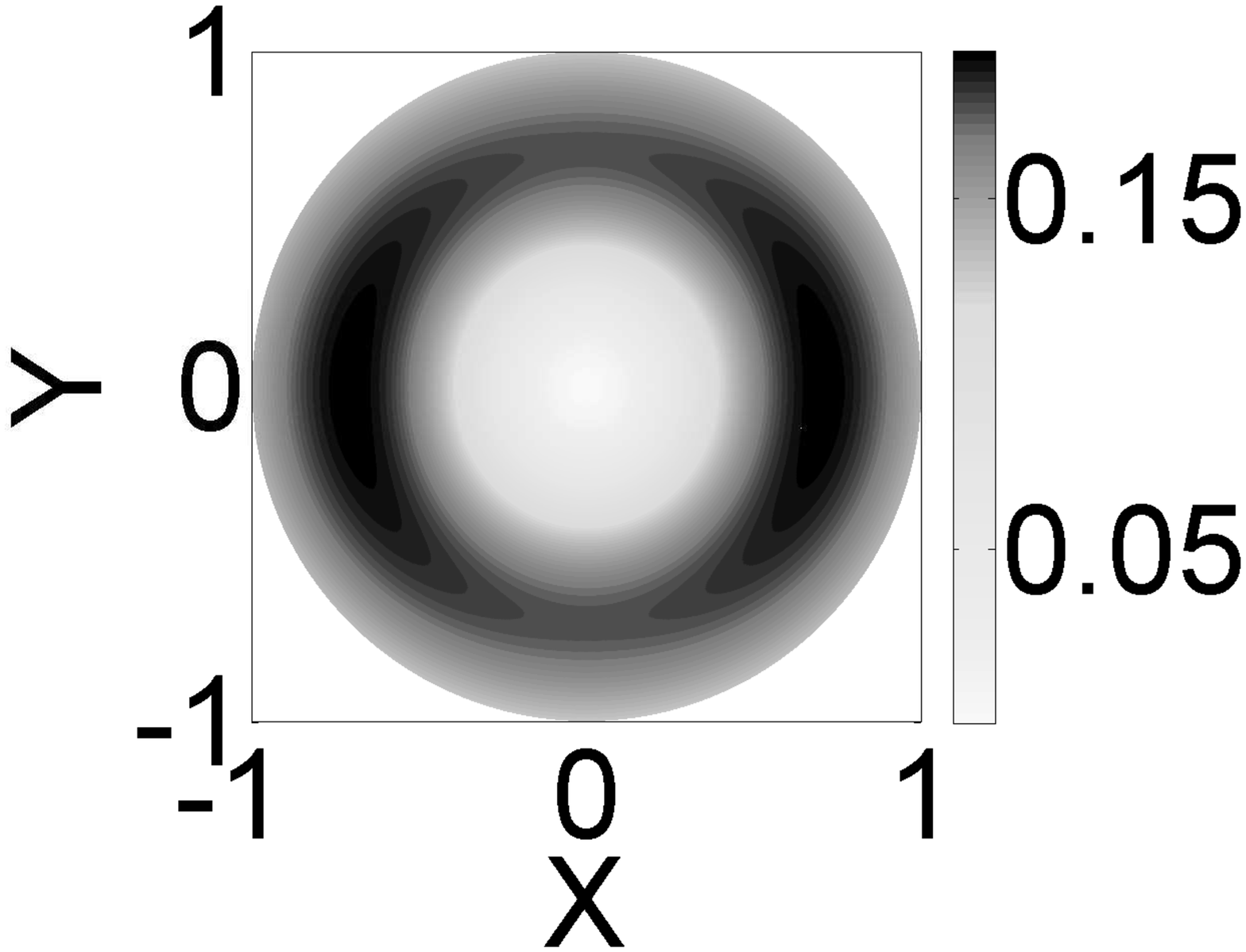}}\\
\subfloat[]{\includegraphics[trim = 1cm 1.5cm 1cm 0cm,scale=.22]{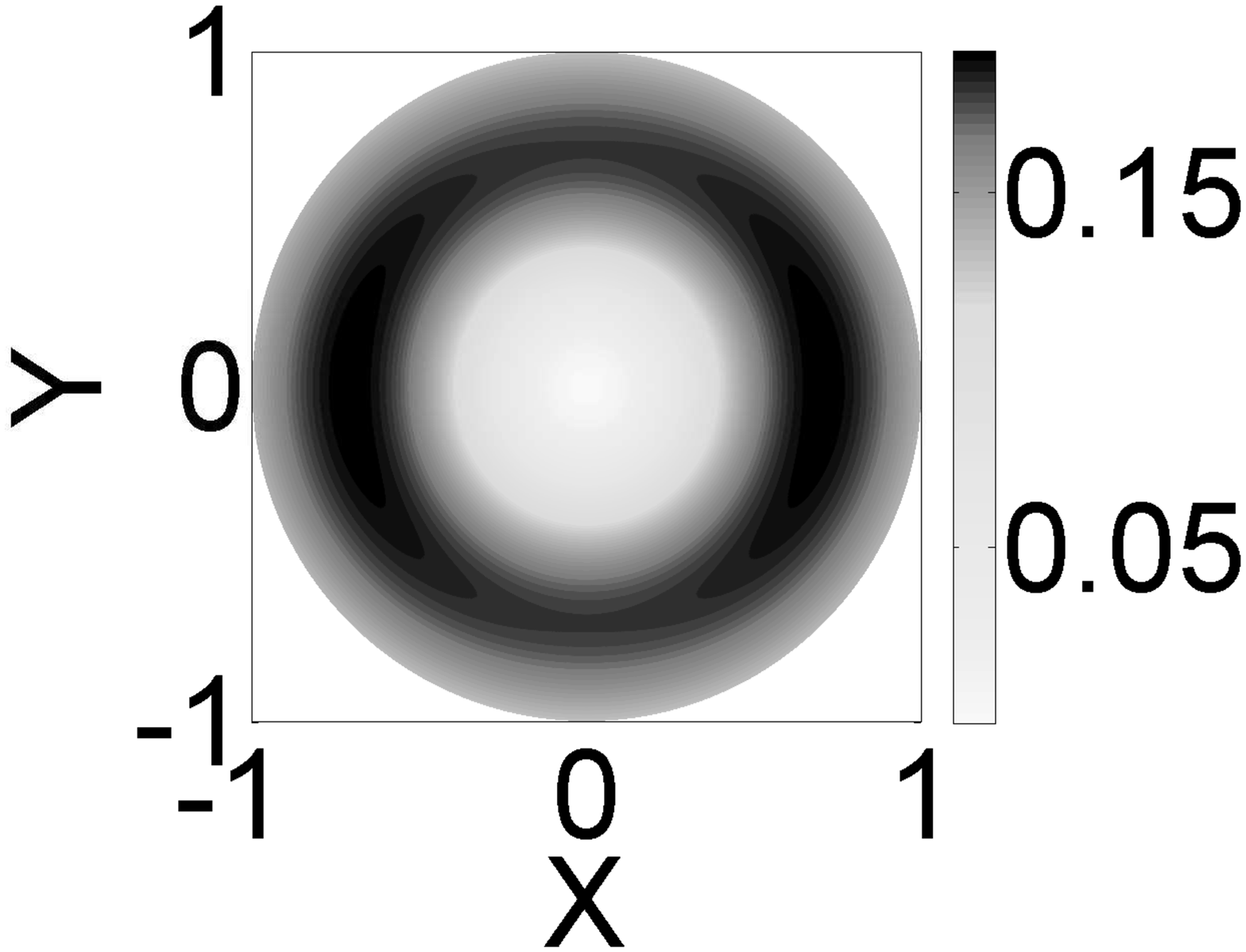}}
\subfloat[]{\includegraphics[trim = 1cm 1.5cm 1cm 2cm, scale=.22]{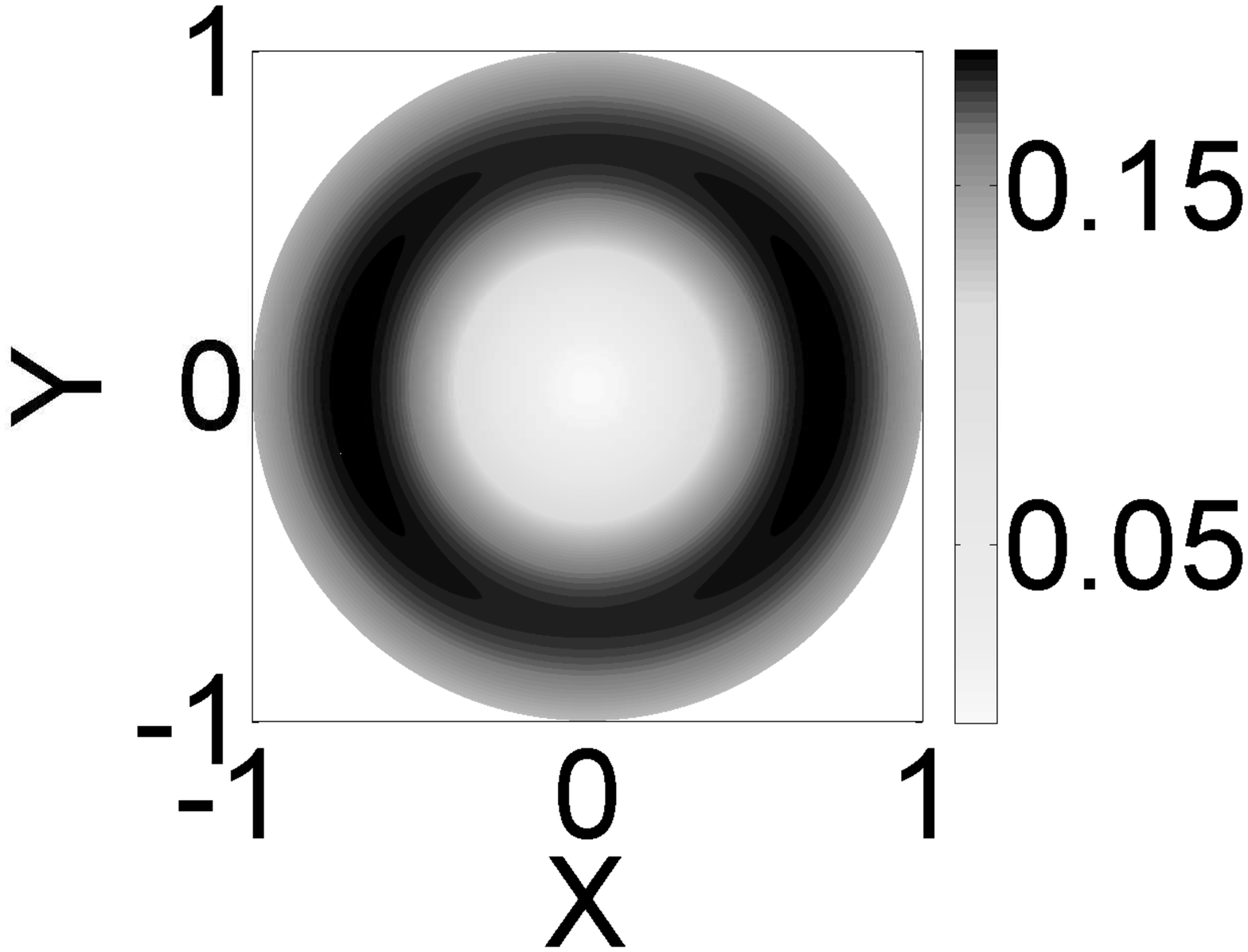}}
\caption{Plot of  the density distribution of  vortex anti-vortex states  for focusing angles (a) 70$^\circ$,  (b) 60$^\circ$, (c) 50$^\circ$, (d) 40$^\circ$.  All quantities are in dimensionless units.}
\end{figure*}

In TABLE II. the Rabi frequencies are calculated, considering OAM=+1 and SAM=+1 of paraxial field. Here, the three channels with different intermediate states are  $\Omega_4=  \lvert F^\prime=2, m_f =0 \rangle $, $\Omega_5=  \lvert F^\prime=2, m_f =-2 \rangle$ and $\Omega_6=  \lvert F^\prime=2, m_f =-1 \rangle$ with vorticities 1, 3, 2 respectively. Therefore, a superposition of these three  vortex states is possible with comparable combination from each of them. At high focusing angle, vortex states correspond to $\kappa=$2 and 3 dominate over $\kappa=$1, which is the only possible vortex state when the LG beam is non-focused. Also, TABLE II shows that, at higher focusing angle the Rabi frequency $\Omega_5$ dominates over  the same to  $\Omega_6$ and the crossover between the two frequencies  takes place at focusing angle $\approx 20^\circ$.

\begin{table}[h]
\caption{Magnitude of Rabi frequencies (in MHz)  of two-photon Raman transitions for different focusing angle of incident beam of OAM=+1, SAM=+1. $\kappa$ is the final vorticity of atoms in BEC.} % title of Table
\centering % used for centering table
\begin{tabular}{c |c|c | c} % centered columns (4 columns)
\hline\hline %inserts double horizontal lines
Focusing angle  & $\Omega_4 (\kappa=1)$ & $\Omega_5 (\kappa=3)$ & $\Omega_6 (\kappa=2)$ \\ [0.2ex] % inserts table
%heading
\hline % inserts single horizontal line
70$^\circ$ & $7.61$ & $6003.60 $ & $367.31 $ \\ % inserting body of the table
60$^\circ$ & $6.44$ & $3302.00 $ & $248.12 $ \\
50$^\circ$ & $5.06$ & $1370.10 $ & $144.74 $ \\
40$^\circ$ & $3.59$ & $433.79 $ & $69.33$ \\
30$^\circ$ & $2.19$ & $89.26$ & $24.33$ \\ 
20$^\circ$ & $1.03$ & $8.69$ & $5.27$ \\
10$^\circ$ & $0.27$ & $0.14$ & $0.35$ \\[0.2ex]
 % [1ex] adds vertical space
\hline %inserts single line
\end{tabular}
\label{table:nonlin} % is used to refer this table in the text
\end{table}

\section{CONCLUSION}

We have developed the theory of interaction of non-paraxial LG beam with matter. Since, OAM and SAM are no longer conserved separately, the interaction can take place through three different orbital angular momentum channels.  Therefore, the total angular momentum of optical beam is distributed among the c.m. and electronic motions of atoms in three possible ways.  We have prescribed a possible method of creating of superposition of vortex states using  single  LG pulse and three Gaussian pulses using two-photon stimulated Raman transition. Our numerical calculations estimate   the variation of number of atoms in different vortex states with the focusing angle. At high focusing angle, we see the possibility of interference pattern created from vortex and anti-vortex.  As we have gone beyond paraxial limit, many new properties of interaction have been emerged which can have profound applications in different areas of science and technology in future.

\appendix*
\section{} 

Interaction Hamiltonian derived in the (PZW) scheme
\begin{equation}
H_{int}=-\int d\textbf{r}^\prime P(\textbf{r}^\prime)\boldsymbol{.} \textbf{E}(\textbf{r}^\prime, t) +H.c.
\end{equation}
where $P(\textbf{r}^\prime)$ is the electric polarization
given by
\begin{equation}
P(\textbf{r}^\prime)=-e\dfrac{m_c}{m_t}\textbf{r}\int_0^1 d\lambda \delta \Big(\textbf{r}^\prime-\textbf{R}-\lambda\dfrac{m_c}{m_t}\textbf{r}\Big).
\end{equation}
We  use the Taylor's expansion for the electric field about $\textbf{R}$
\begin{equation}
E_i\Big(\textbf{R}+\lambda \dfrac{m_c}{m_t}\textbf{r}\Big)=E_i(\textbf{R})+ \lambda \dfrac{m_c}{m_t}  [{\vec{r}}. \vec{\nabla} E_i(\textbf{r})]_\textbf{R} +. . .
\end{equation}

Here $i$ refers to the  x, y, and z component of the electric field. We will use the 1st part of the Taylor's expansion to determine the electric dipole transition. 2nd part shows the effect of  electric field gradient which finally estimate the electric quadrupole transition.  Using eq. (A.1 - A.3), the interaction Hamiltonian can be written as,

\begin{equation}
H_{int}= e\dfrac{m_c}{m_t} \textbf{r} \boldsymbol{.} \textbf{E}\Big(\textbf{R}+\lambda \dfrac{m_c}{m_t}\textbf{r}\Big)
\end{equation}

If, we are focusing only on the  electric dipole transition,

\begin{eqnarray}
H_{int}&=&  e\dfrac{m_c}{m_t} E_0 \textbf{r}\boldsymbol{.}\Bigl[(-i)^{l+1} I_0^{(l)}(R_\bot,Z)e^{il\Phi}\boldsymbol{\hat{\textbf{x}}}+(-i)^{l+1} I_{2\beta}^{(l)}(R_\bot,Z)e^{i(l+2\beta)\Phi}\boldsymbol{\hat{\textbf{x}}} \nonumber \\ &+& \beta(-i)^l  I_0^{(l)}(R_\bot,Z)e^{il\Phi} \boldsymbol{{\hat{\textbf{y}}}}- \beta(-i)^l I_{2\beta}^{(l)}(R_\bot,Z)e^{i(l+2\beta)\Phi}{{\hat{\textbf{y}}}}\nonumber \\
&-&(2\beta) (-i)^{l} I_\beta^{(l)}(R_\bot,Z)e^{i(l+\beta)\Phi}\boldsymbol{\hat{\textbf{z}}} \Bigr].
\end{eqnarray}
Here, we used the expression of electric field components from eq. (1 - 3)  to determine  eq. (A.5). 
After rearranging this  equation, 

\begin{eqnarray}
H_{int}&=&  e\dfrac{m_c}{m_t} E_0 \textbf{r}\boldsymbol{.}\Bigl[I_0^{(l)}(R_\bot,Z)e^{il\Phi}\{\boldsymbol{\hat{\textbf{x}}}(-i)^{l+1}+ \boldsymbol{{\hat{\textbf{y}}}} \beta(-i)^l\} \nonumber \\
&+& I_{2\beta}^{(l)}(R_\bot,Z)e^{i(l+2\beta)\Phi}\{\boldsymbol{\hat{\textbf{x}}}(-i)^{l+1}  -\boldsymbol{{\hat{\textbf{y}}}}\beta(-i)^l\}\, \nonumber\\
&-&(2\beta) (-i)^{l} I_\beta^{(l)}(R_\bot,Z)e^{i(l+\beta)\Phi}\boldsymbol{\hat{\textbf{z}}} \Bigr].
\end{eqnarray}

Now for $l=+1$ and $\beta=\pm 1$, the Hamiltonian has the form,
\begin{eqnarray}
H_{int}^{ l=+1,\beta=\pm 1}&=& e\dfrac{m_c}{m_t} E_0 \textbf{r}\boldsymbol{.}\Bigl[-I_0^{(1)}(R_\bot,Z)e^{i\Phi}\{\boldsymbol{\hat{\textbf{x}}}\pm i \boldsymbol{{\hat{\textbf{y}}}} \} \nonumber \\
&-& I_{\pm 2}^{(1)}(R_\bot,Z)e^{i(1\pm 2)\Phi}\{\boldsymbol{\hat{\textbf{x}}}\mp i \boldsymbol{{\hat{\textbf{y}}}} \}\, \nonumber\\
&\pm & \sqrt{2} i I_{\pm 1} ^{(1)}(R_\bot,Z)e^{i(1 \pm 1)\Phi}\boldsymbol{\hat{\textbf{z}}} \Bigr].
\end{eqnarray}

 Using the condition $\textbf{r . E}_0=r \sqrt{\frac{4\pi}{3}}\sum_{\delta=0,\pm 1}\epsilon_{\delta}Y_1^{\delta}(\boldsymbol{\hat{\textbf{r}}}),$ with $\epsilon_{\pm 1}=(E_x \pm i E_y)/\sqrt{2}$ and $\epsilon_0 = E_z$, we get
\begin{eqnarray}
H_{int}^{ l=+1,\beta=\pm 1}&=&  e\dfrac{m_c}{m_t} r \sqrt{\dfrac{8\pi}{3}}\Bigl[-I_0^{(1)}(R_\bot,Z)e^{i\Phi}\epsilon_{\pm 1}Y_1^{\pm 1}(\boldsymbol{\hat{\textbf{r}}})\nonumber \\
&-& I_{\pm 2}^{(1)}(R_\bot,Z)e^{i(1\pm 2)\Phi}\epsilon_{\mp 1}Y_1^{\mp 1}(\boldsymbol{\hat{\textbf{r}}})\, \nonumber\\
&\pm & \sqrt{2} i I_{\pm 1} ^{(1)}(R_\bot,Z)e^{i(1 \pm 1)\Phi}\epsilon_{=0}Y_1^{0}{(\boldsymbol{\hat{\textbf{r}}}})\Bigr].
\end{eqnarray}
%) ============================================================================
% === REFERENCES =============================================================
% ============================================================================

%\bibliographystyle{abbrv}

%\bibliography{bib}\

\end{document}